# Quantitative Raman Spectrum and Reliable Thickness Identification for Atomic Layers on Insulating Substrates

Song-Lin Li,[1,2,*] Hisao Miyazaki,[1] Haisheng Song,[1] Hiromi Kuramochi,[1] Shu Nakaharai,[3] and Kazuhito Tsukagoshi[1,*]

[1]WPI Center for Materials Nanoarchitechtonics (WPI-MANA) and [2]International Center for Young Scientist (ICYS), National Institute for Materials Science, Tsukuba, Ibaraki 305-0044, Japan
[3]Collaborative Research Team Green Nanoelectronics Center, National Institute of Advanced Industrial Science and Technology, Tsukuba, Ibaraki 305-8569, Japan

E-mail: li.songlin@nims.go.jp and tsukagoshi.kazuhito@nims.go.jp

**Abstract**

We demonstrate the possibility in quantifying the Raman intensities for both specimen and substrate layers in a common stacked experimental configuration and, consequently, propose a general and rapid thickness identification technique for atomic-scale layers on dielectric substrates. Unprecedentedly wide-range Raman data for atomically flat $MoS_2$ flakes are collected to compare with theoretical models. We reveal that all intensity features can be accurately captured when including optical interference effect. Surprisingly, we find that even freely suspended chalcogenide few-layer flakes have a stronger Raman response than that from bulk phase. Importantly, despite the oscillating intensity of specimen spectrum *versus* thickness, the substrate weighted spectral intensity becomes monotonic. Combined with its sensitivity to specimen thickness, we suggest this quantity can be used to rapidly determine the accurate thickness for atomic layers.

Atomically thin two-dimensional (2D) crystals, including graphene,[1] exfoliated chalcogenides,[2,3] self-organized nanosheets[4,5] and topological insulators,[6,7] have generated intensive research due to their scientific significance and technological potential. Stemming from drastic dimension reduction, numerous intriguing phenomena are observed, such as Dirac dispersion relation,[1,6,7] variable band structure[8-10] and helical Dirac fermions.[6] For these materials, interesting phenomena are not necessarily limited in monolayer flakes and, sometimes, attractive properties emerge in samples with increased thickness. For instance, the surface state induced Dirac cones develop only in $Bi_2Se_3$ films thicker than five quintuple layers,[7] while under perpendicular electric field energy gaps form in bilayer rather than monolayer graphene.[11] Accurate thickness information and sufficient characterization range is particularly important for the ultrathin materials. So far, however, there have been few rapid and nondestructive thickness characterization techniques for the inorganic atomic layers. A direct transfer of the well established schemes from graphene, such as optical contrast[12,13] and Raman phonon position,[14,15] to the inorganic flakes seems not so successful. For example, the optical contrast exhibits a nonmonotonic response and a low sensitivity for few-layer chalcogenide flakes in most excitation wavelengths[12] which largely reduces the convenience of use. Developing a general and effective thickness characterization scheme represents a strong desire from the scientific community.

On the other hand, it is well recognized that the optical interference has a strong impact on the intensity of Raman spectrum.[16-19] This phenomenon draws renewed attention[20-22] after the isolation of 2D graphene in 2004. In an important advance, Wang *et al.* first point out that the multiple reflections within the graphene and dielectric layers are responsible for the strong modulated Raman response at varied graphene thicknesses.[20] However, a discrepancy still exists between experiment and calculation, which could be understood as experimental errors due to inevitable interfacial roughness before 2004, but is hard to accept presently when atomically flat graphene flakes are used. This leaves a doubt whether there are other factors, such as surface plasmon,[23] involved in the Raman process. To solve this issue and strictly verify the interference effect on Raman spectrum, an independent study on other materials is highly desired.

Propelled by the two motivations above, we performed an unprecedented Raman investigation on atomically flat MoS$_2$ flakes over a wide range from 1 to ~120 consecutive layers. We demonstrated that optical interference is the dominant factor affecting spectral response and managed to quantify the Raman intensities for both the MoS$_2$ specimen and Si substrate layers in the common stacked sample configuration. Excellent agreements between the interference-based models[21,22] and calculated data were achieved. In addition, by extracting the ratio of spectral intensity of MoS$_2$ to Si, we showed that this intensity ratio is a monotonic spectral invariant *versus* specimen thickness and is capable of identifying MoS$_2$ thickness. By appropriately incorporating the interference effect, strong responses (2–20 folds with respect to bulks) can be achieved from the atomic layers, no matter freely suspended or supported by substrates. Raman spectra for other chalcogenides were also calculated and showed similar behavior as MoS$_2$, indicating a good generalization of above results. These results enable a quantitative understanding on Raman spectroscopy that may lead to versatile applications, such as rational design for Raman enhancement and thickness characterization for ultrathin structures.

**Results and Discussion**

The chalcogenide MoS$_2$, an important electronic material—next generation graphene,[24,25] is chosen as the specimen because of its marked cleavage properties and potential applications in short-channel transistors[2,26-28] and optoelectronic components.[29] Interestingly, when thinned from bulk to monolayer, its band structure undergoes an indirect to direct change and the photoluminescence efficiency increases accordingly.[8-10] Besides, the energy gap and phonon modes also depend on the number of layers (NL).[14] The structure of the 2H-MoS$_2$ chalcogenides (space group: P6$_3$/mmc) is illustrated in Figure 1a, in which one cation (Mo$^{+4}$) plane is sandwiched between two anion (S$^{-2}$) planes and the layered structure arises from the stacking of hexagonally packed sheets in sequence.[30] Figures 1b and 1c show optical and atomic force microscopy (AFM) images for MoS$_2$ flakes from consecutive 1 to 4 layers. In our AFM measurements, the tapping mode was employed to minimize sample damage. Figure 1d shows the average profile of the rectangular area presented in Figure 1c. The linear fit of the layer heights (Figure 1e) reveals a fake height of ~ 1.3 nm for bare SiO$_2$/Si substrates, which may arise from the different tapping responses of substrate and MoS$_2$, and a step height of 0.70 nm between consecutive layers, which is slightly larger than the previous report of 0.615 nm.[30] The NL values from 1 to 3 was determined from the distance of the Raman modes of $E_{2g}^1$ and $A_{1g}$ (Figure 2e) which is more accurate than AFM, while the values for NL>3 are assigned through combined AFM measurement and optical contrast which brings out 10% error in the "nominal" NL values of our flakes. Raman spectra were taken for consecutive MoS$_2$ flakes and over a wide NL range from 1 to ~ 120 layers.

One of the reasons for extending Raman measurements to thick flakes was to determine the critical thickness for dimensionality crossover from 3D to 2D. Figure 2a shows typical Raman spectra for different NL values, and the spectral information (peak position, area, height and width) is extracted with Lorentzian fittings. Two sharp Raman modes $E_{2g}^1$ (~383 cm$^{-1}$) and $A_{1g}$ (~ 408 cm$^{-1}$) modes are observed and exhibit strong NL dependence. The first decreases from 386 to 383 cm$^{-1}$ and the second increases from 404 to 408 cm$^{-1}$ as NL increases from 1 to 20 (Figures 2b and 2c), consistent with the observation of Lee *et al.*[14] These shifts are attributed to the variation in the dielectric screening environment for long-range Coulomb interactions as NL changes.[31] In addition to the above first-order modes, a rather weak second-order scattering process, 2×LA(*M*) mode,[32] near 452 cm$^{-1}$ was recorded. This mode is also NL dependent, increasing from 447 to 452 cm$^{-1}$. Figures 2b–2e reveal that the 3D properties for the lattice phonon modes of MoS$_2$ persist to a thickness at least 10 layers and the 2D properties become essential as NL < 5.

An important finding here is the observation of an interference induced high-order Raman enhancement peak, which was never seen in analogue systems.[14,20] Due to the narrow NL range covered in previous studies, only one enhancement peak is observed at NL ~ 10 in graphene[20] and at NL ~ 4 in MoS$_2$.[14] It is generally believed that no additional local maxima exist in thicker layers.[20] However, we identified a new enhancement peak at NL ~ 85 in MoS$_2$. In Figures 2f and 2g, the multiple enhancement peaks manifested themselves as two intensity maxima when plotting peak area and height as a function of NL. All three Raman modes exhibit the same intensity tendency, indicating that this enhancement effect is independent of the lattice vibration modes. The colors of the MoS$_2$ flakes are also suggestive of the interference effect. Under white light illumination, the flakes with thickness near the first and second peaks are dark blue and pink, respectively, while the flake with a thickness around the first valley exhibits a dim white color (inset of Figure 2f).

As mentioned, although efforts have been made in understanding the spectral response as a function of graphene thickness,[20] a large discrepancy remains between the calculation and experiment. The most important contribution here is accurately quantifying the Raman spectra over a wide NL range, which allows us to rule out the possibility of other factors, such as surface plasmon,[23] engaging in the Raman process so that we can draw an affirmative conclusion that optical interference is the sole modulation source. Note that strict optical relations for three-layer systems are quite complicated.[18,19] Similar to previous works,[20,21] a simplification made here is only the normal incidence is considered so that the p- and s-components of excitation can share the same expression. We also checked that such a simplification would not cause large deviation, because most additional contributions due to oblique incidence cancel out between p- and s-components and the majority of light is close to normal incidence condition due to the Gaussian distribution of laser energy (Section 3,

Supporting Information). As will be seen below, this first-order approximation catches the main experimental features and gives a satisfied accuracy to fit with experiment.

The optical paths for the excitation and scattering light are quite complex because the incident light undergoes an infinite number of reflections and refractions at the boundaries of both MoS$_2$ and SiO$_2$ layers (Figure 3a). A strategy for solving this optical issue is to first calculate the effective reflection coefficient at the MoS$_2$/SiO$_2$ interface by accounting for multiple reflections in the SiO$_2$ dielectric layer, and then analyze the light distribution in the MoS$_2$ specimen layer.[20] For convenience, the four involved media are designated by the index $i$, and the corresponding complex refractive indices are represented by $\tilde{n}_i$, where $i = 0, 1, 2$ and 3 for air, MoS$_2$, SiO$_2$ and Si, respectively. After including the multiple reflections, the output Raman intensity from the top MoS$_2$ layer (total thickness $d_1$) can be expressed as[21]

$$I = \int_0^{d_1} |F_{ex}(x)F_{sc}(x)|^2 dx, \quad (1)$$

where $F_{ex}(x)$ and $F_{sc}(x)$ are the electric field amplitudes for the excitation and scattering light, respectively. The derivation and the full expressions of them are given in Section 1.1–1.3 in the Supporting Information.

Figure 3b compares the calculation and experiment for the $E_{2g}^1$ mode at varied NL values. The calculation agrees well with the experiment in terms of the peak positions and the spectral intensity from 4 to 120 layers. For instance, it duplicates the two peak positions at NL ~ 4 and 80 and their intensity ratio ~ 3. Such an excellent agreement is rather surprising since the current calculation contains no fitting parameters. This agreement also unambiguously indicates the exclusive modulation role played by optical interference in the stacked systems. For NL < 4, reduced Raman responses are observed, which can be attributed to the decreased real thickness of few-layer flakes compared with the theoretical values we adopted in calculation by using the integer times of layer spacing in bulk. The phenomenon of thickness reduction is common in ultrathin materials and was reported in graphene and nanotube systems.[33] Extending the calculation to large NL regime reveals the existence of 4 enhancement peaks within 300 layers. Their enhancement factors (relative to bulks) decay from 10, 3.0 and 1.3 to 1.1. For NL > 300, no clear enhancement peak exists. In addition to the enhancement peaks from constructive interference, valleys due to destructive interference are also observed. The intensity of the first valley at NL ~ 45 is only half of the bulk value. The presence of both constructive peaks and destructive valleys further confirms the interferential nature of the observed spectra.

The accurate control on the specimen thickness also enables an interesting observation on the Raman spectrum of substrate layer. Actually, in the $MoS_2/SiO_2/Si$ stack the response from the Si substrate at 520 cm$^{-1}$ is not only rather strong but also close to the two $MoS_2$ main modes of $E_{2g}^1$ and $A_{1g}$ (Figure 2a). The three peaks are inevitably recorded together during collection. An analysis on the 'byproduct' of Si peak also helps to check the validity of established interference model. Here we find that besides the specimen spectrum, the interference model describes the substrate spectrum as well. The related derivation and expression for the Si spectrum can be found in Section 1.4 in the Supporting Information. In Figure 3d, the calculated spectrum *versus* NL of $MoS_2$ is plotted and characterized by a dominant exponential decay, which results from the strong absorption of incident light by the $MoS_2$ layer above $SiO_2$. The interferential feature from the Si spectrum is not as appreciable as the $MoS_2$ spectrum, but still discernable at NL ~ 80 when plotted the intensity logarithmically in Figure 3c. The successful duplication of the weak fine structures confirms again the validity of established interference models.

A surprising finding in this work is that a $MoS_2$ monolayer, no matter freely suspended or placed above $SiO_2$, can have stronger Raman response than bulks. This strikingly contradicts the intuition that atomic layers would have much weak signals due to the drastic amount reduction. Figure 3e shows the enhancement factor for the suspended and supported $MoS_2$ flakes with respect to bulks. We choose the bulk phase as reference because such a configuration excludes all interference paths and corresponding spectrum is easy to obtain. For a monolayer, the enhancement factors reach 2.5 and 6 in the suspended and supported configurations, respectively. The highest enhancement factor for freely suspended MoS$_2$ flaks is 5 at NL ~ 4, while the value doubles when an additional 285 nm SiO$_2$ layer is employed as an interference enhancement layer.

To understand this phenomenon, the intensity distributions of excitation light within the MoS$_2$ flakes are calculated under different specimen configurations. For the bulk configuration, the light follows a traditional light absorption process, that is, an exponential decay from incident position. The initial intensity relates to transmittance coefficient $t_{01}$ (from air to MoS$_2$) and is calculated to be ~$0.5|E_0|^2$, where $|E_0|^2$ is the intensity of excitaion laser. For freely suspended and supported 5 layers, the intensities are almost fixed at ~$3|E_0|^2$ and $4|E_0|^2$, respectively, both higher than that in the bulk configuration (Figure 3f). This theoretical result thus provides a fundamental support for investigating the intrinsic spectral behavior for freely suspended samples.[34] Additionally, the light distributions in the suspended and supported configurations are highly dependent on NL, controlled by the two processes of optical absorption and interference (Figures 3g–3i). When NL > 200, the distributions in the two configurations approach that in the bulk.

Another essential motivation of this work is to develop a general and rapid criterion for counting NL for atomic inorganic flakes. The full quantification on spectral behavior enables us to reach the goal by using the intensity ratio of MoS$_2$ to Si as the criterion (Figure 4), as did in graphene.[22] In calculating the intensity ratios, the scattering cross sections for the MoS$_2$ $E_{2g}^1$ and A$_{1g}$ modes were taken as 2.3 and 3.9 times of that of Si substrate, respectively. It is evident that despite the oscillating intensity of the MoS$_2$ spectrum (Figure 2f), the weighted intensity by Si spectrum becomes monotonic for all NL range (inset of Figure 4), making it rival the previous optical contrast method.[12,13] As shown in Figures 4a and 4b, both the $E^1_{2g}$ and $A_{1g}$ modes can be used and the most sensitive range spans from 1 to 20 layers. This new criterion, in principle, covers all NL range. Its limitation to large NL regime stems from the fast decay in the Si spectrum and the increasing fitting uncertainties (Figure 3d). For less absorbed specimens, the detection range is expected to extend. Error analysis is also performed for this identification method and shown in Figures 4c and 4d. For NL ≤ 7 the intensity ratios are discrete enough to discern each NL values, while the error is one layer in the 7 < NL < 15 regime and increases to two layers in the NL > 15 regime. The overall error is concluded to be ±10 % for the investigated range. Nevertheless, the 20 layer detection ability and ±10 % thickness accuracy are sufficient in most cases for the low-dimensional studies on atomic layers. The intensity ratio for varied $SiO_2$ thicknesses is also calculated and given in Figures 4e and 4f. Sufficient detection resolutions are disclosed when the $SiO_2$ thickness is changed by ±30 nm around optimal values of 91 and 273 nm. It deserves noting that to achieve excellent identification resolutions some specific $SiO_2$ thickness ranges that causes destructive interference should be avoided, as will be discussed later.

To fully understand the spectrum for rational designs for Raman enhancement, we further calculated the dependence of the spectral intensity on three main experimental factors: the NL of $MoS_2$ ($NL_{MoS_2}$), the $SiO_2$ thickness ($d_2$) and the excitation wavelength ($\lambda_{ex}$). Figure 5a shows a contour plot of the enhancement factor (excited at 532 nm) as a function of $NL_{MoS_2}$ and $d_2$. The irregular traces of the constructive peaks (dotted lines) are characteristic of the roles played by $NL_{MoS_2}$ and $d_2$ in the interference phase factor $\phi = 2\pi\tilde{n}_1 d_1 / \lambda_{ex} + 2\pi\tilde{n}_2 d_2 / \lambda_{ex}$. When $NL_{MoS_2}$ ($\propto d_1$) increases, $d_2$ has to decrease to maintain the constructive condition, $\phi = (N + 1/2)\pi$ ($N$ = integer). This provides a basic reference in optimizing dielectric thickness for detecting atomic layers. As far as monolayer are concerned, the optimal $SiO_2$ thickness is $d_2 \sim (2N+1)\lambda_{ex}/4n_2$. An enlarged plot for the ultrathin range (1–20 layers) is given in Figure 5b. The horizontal axis is reduced to the period factor $2n_2d_2/\lambda_{ex}$ to eliminate explicit experimental parameters. The maximum enhancement factors range from 3 to 14 (inset of Figure 5b). It is also important to keep in mind the existence of destructive interference when inappropriate dielectric thicknesses are used, which would lead to a reduction in spectral intensity by 1–2 orders of magnitude (Figure 5c). Therefore, a careful thickness arrangement on specimen and dielectric is necessary.

When calculating the wavelength dependence, special attention was paid to the dispersion of refractive indices ($\tilde{n} = n - ik$) with wavelength, as well as the well-known quartic dependence of scattering cross section on excitation frequency $\sigma \propto f^4$. The $n$ and $k$ values of MoS$_2$ (and other four chalcogenides) from 300 to 900 nm excitation are explicitly given in Figure 5d (and Figures S4–S7), which may be useful for future studies. The enhancement factor exhibits oscillating patterns with respect to the phase factor $2n_2d_2/\lambda_{ex}$. Its magnitude is highly dependent on the combination of $n$ and $k$ values, with the highest value reaching 22 near 490 nm (Figure 5e). At $\lambda_{ex} > 700$ nm the factor approaches zero, resulting from the largely increased bulk response due to the reduction in light absorption in long-wavelength regime ($k \sim 0$, Figure 5d). High raw intensity is located in high-frequency regime due to the quartic dependence of scattering cross section on frequency (Figure 5f), indicating that high-frequency excitation helps to obtain strong response.

Additional efforts were made to calculate the spectra for other four chalcogenides, MoSe$_2$, MoTe$_2$, WS$_2$ and WSe$_2$ (Figures S4–S7) since they may contribute to ultrathin-channel electronics as MoS$_2$. The NL dependent enhancement factor for the $E_{2g}^1$ modes are shown in Figure 6. In low NL regime, all the five materials have a stronger response than corresponding bulks with enhancement factor from 2 to 15, which suggests that a sole interference enhancement is enough to achieve sufficient Raman signals for these atomic layers. Second-order enhancement peaks appear in all materials, but with distributed intensities. The position of the second enhancement peak follows the sequence $MoS_2$ < $MoSe_2$ < $WS_2$ < $MoTe_2$ < $WSe_2$, in line with the magnitude of the real part *n* in their refractive indices (Table S2, Supporting Information). Normally, high-order peaks are strong in materials with small *k* (imaginary part of refractive index) and large *n* values, such as $WS_2$ and $MoS_2$. This is because a small *k* results in low sample absorption and large interferential components, and a large *n* leads to a short optical path required for interference and thus reduces absorption. Both factors are beneficial for light interference and final peak intensity. This understanding enables to recheck the situation in graphene which has a refractive index around 2.66-1.33i, being small *n* and large *k* values as compared with chalcogenides. Therefore, its high-order enhancement peaks are not strong (Figure S2, Supporting Information) and tend to be hidden in noise during measurements. This result explains why multiple enhancement peaks are hard to be observed in graphene.[20]

**Conclusion**

We have conducted extensive measurements and calculations on the Raman spectra of chalcogenide flakes on dielectric substrates. For the first time, we observe clear high-order enhancement peaks in atomically flat samples and reveal the decisive role played by optical interference in the spectra of stacked systems. Impressively, quantitative Raman spectra are achieved in a wide range for both the specimen and substrate layers. We also reveal that even freely suspended few-layer flakes can have stronger response than bulks due to inner optical interference. Besides rational designs for Raman enhancement, we also lay an important theoretical foundation for a thickness identification technique for inorganic atomic layers. The results provide insightful view in the Raman behavior of common stacked systems and would lead to versatile applications.

**Experimental section**

$MoS_2$ flakes were prepared by micromechanical cleavage from commercial $MoS_2$ crystals (Furuchi, Japan) and were transferred to Si wafers with a 285-nm $SiO_2$ capping layer. Hybrid techniques of Raman peak position, AFM and optical contrast spectra were used to determine the NL values for $MoS_2$ flakes. Raman spectra were acquired at an excitation wavelength of 532 nm and a laser power of less than 0.1 mW to avoid sample heating or oxidation in air. An integration time of 30 s was used to enhance the signal-to-noise ratio. The laser beam was focused onto the $MoS_2$ sample by a 100× objective lens with an NA of 0.9. The scattered light was collected and collimated by the same lens. The scattered signal was dispersed by a spectrometer working at 1800 grooves/mm and was detected by a thermoelectrically cooled CCD (charge-coupled device) detector at -60°C. The spectral resolution was 0.7 $cm^{-1}$. All of the Raman spectra were recorded for the same integration time, laser power and focus status. The size of focused beam was about 1 micron and only flakes larger than 2 microns were used. In order to avoid edge and corner effect, all spectra were collected by carefully focusing the beam spot to one layer without overlapping neighboring layers. However, $MoS_2$ flakes typically have a degraded uniformity and smaller sample area as compared with graphene. The flakes with close NL values are hard to identify under optical microscopy in the NL > 7 regime. Although we try to focus incident laser on uniform $MoS_2$ flakes, the results still contain data from nonuniform flakes. This brings about uncertainties when identifying thick samples. The 520-$cm^{-1}$ Si first-order Raman mode was used for calibration.

In theoretical calculation, the refractive index values of Si and $SiO_2$ were adopted from literature.[35,36] The real ($n$) and imaginary ($k$) parts in refractive index of chalcogenides were translated from the corresponding dielectric permittivity and absorption coefficients,[37,38] and the accurate $\tilde{n}$ values at specific wavelength were obtained by data interpolation.

**Supporting information**

Supporting Information includes the expression derivation of Raman scattering for a trilayer system, the Raman response of graphene/graphite, the $MoS_2$ spectrum with objective lens of varied NA values, the calculated spectra for other chalcogenides, the calculated values of intensity ratios for NL identification, and the refractive indices used in the calculations for the four involved optical media. This material is available free of charge *via* the Internet at http://pubs.acs.org.

**Acknowledgements**

This work was supported in part by a Grant-in-Aid for Scientific Research (No. 21241038) from the Ministry of Education, Culture, Sports, Science and Technology of Japan and by the FIRST Program from the Japan Society for the Promotion of Science.

**Figures**

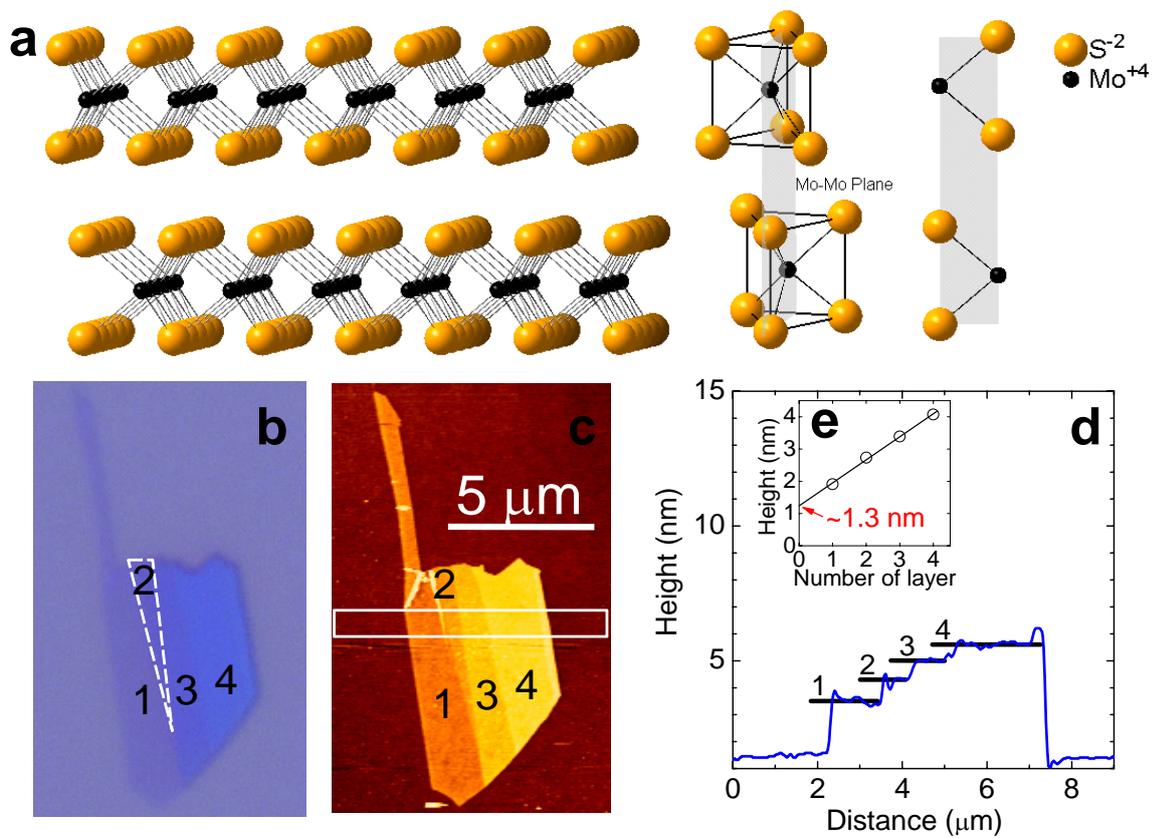

**Figure 1** Atomic Structure and characterization of ultrathin 2H-MoS$_2$ layers. (a) The atomic structure of 2H-MoS$_2$. (b) and (c) Typical optical and AFM images for an exfoliated MoS$_2$ flake with consecutive NL values from 1 to 4. (d) The average height profile for the rectangular area shown in (c). (e) A linear fit of the layer heights from 1 to 4 layers, which gives a base height of ~ 1.3 nm for the bare SiO$_2$/Si substrates and a step height of 0.70 nm between the consecutive layers.

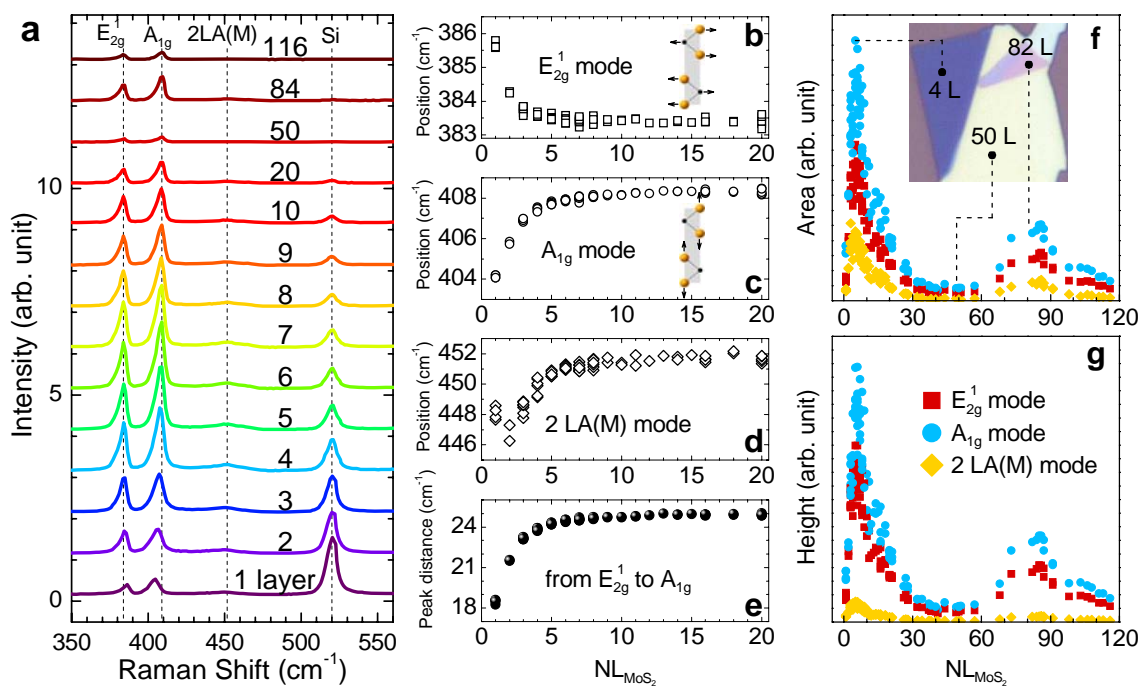

**Figure 2** Raman spectra for MoS$_2$ flakes and evolution of spectral features with thickness. (a) Typical Raman spectra of MoS$_2$ flakes at different NL values from 1 to 116. (b)–(d) Position evolution for the three Raman modes $E_{2g}^1$, A$_{1g}$ and 2×LA(*M*) as a function of NL. (e) The peak distance between the $E_{2g}^1$ and A$_{1g}$ modes. (f) Area and (g) height plots for the three modes as a function of NL. The inset of (f) is an optical image for an MoS$_2$ flake of three typical NL values (4, 50 and 82).

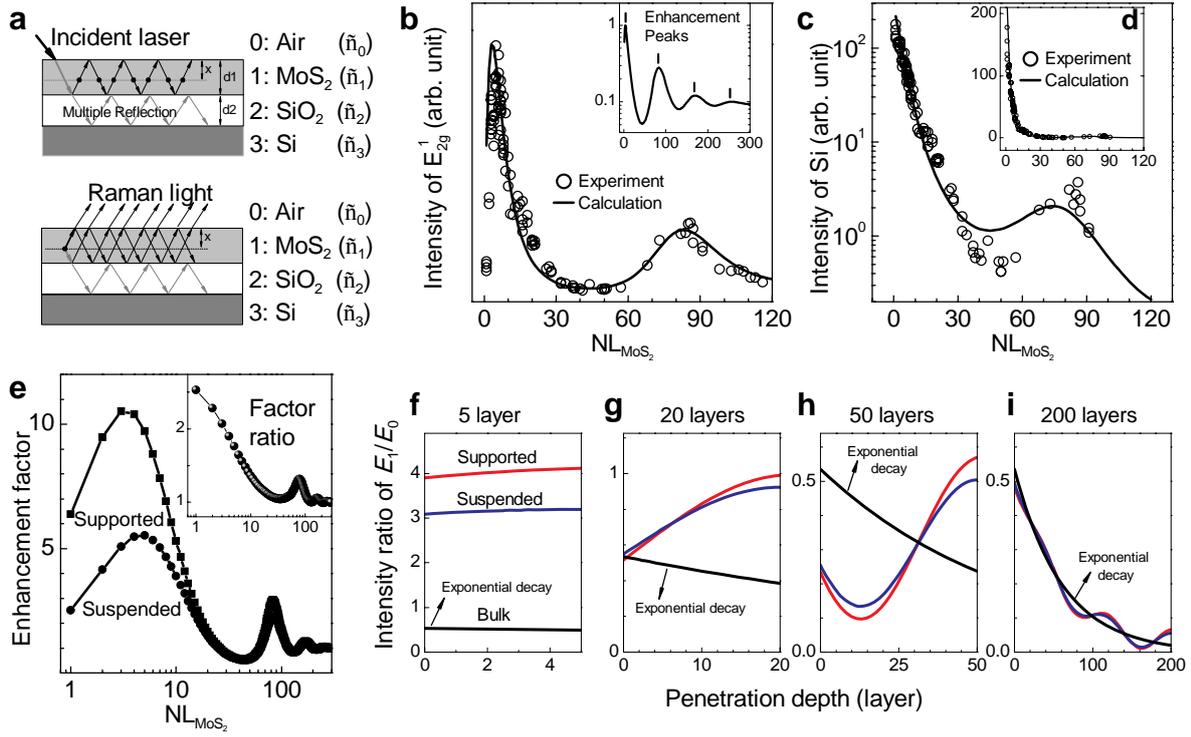

**Figure 3** Modeling and comparison between calculation and experiment for the thickness-dependent spectral intensities in the $MoS_2/SiO_2/Si$ stack. (a) Schematic diagrams for the optical paths of the excitation and Raman scattering light, respectively. (b) The calculation and experiment for the $MoS_2$ layers at different NL values. (c) and (d) The corresponding results for the Si substrate, plotted in logarithmic and linear scales, respectively. (e) Calculated enhancement factor for suspended and supported (on 285 nm $SiO_2$) $MoS_2$ flakes of different thicknesses. The inset is the corresponding factor ratio for the two geometries. (f)–(i) The distribution of excitation light within $MoS_2$ flakes of selected NL values of 5, 20, 50 and 200 layers. For the bulk, the intensity of the excitation light follows a exponential decay with a starting intensity $\sim 0.5|E_0|^2$. The low initial value is due to the low light transmissivity from air to $MoS_2$ ( $|t_{01}|^2 = |2\tilde{n}_0/(\tilde{n}_0 + \tilde{n}_1)|^2 \sim 0.1$ ).

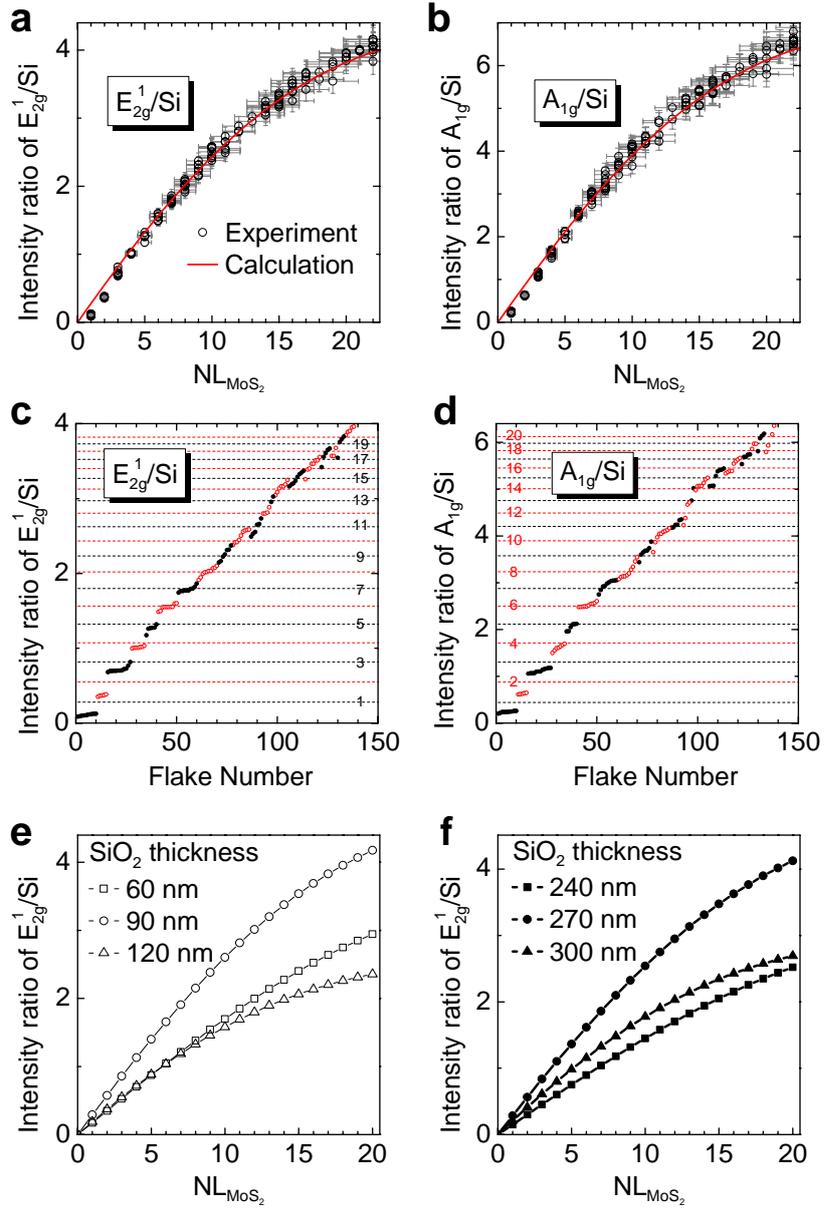

**Figure 4** (a)-(d) Comparison between the calculation and experiment for intensity ratio of the MoS$_2$ $E^1_{2g}$ (~ 383 cm$^{-1}$) and A$_{1g}$ (~ 408 cm$^{-1}$) modes to that of the Si substrate (520 cm$^{-1}$). The errors for the assgined NL values and intensity ratios are 10% and 5%, respectively. (e) and (f) Calculated intensity ratios at different SiO$_2$ thicknesses around optimal values of 91 and 273 nm.

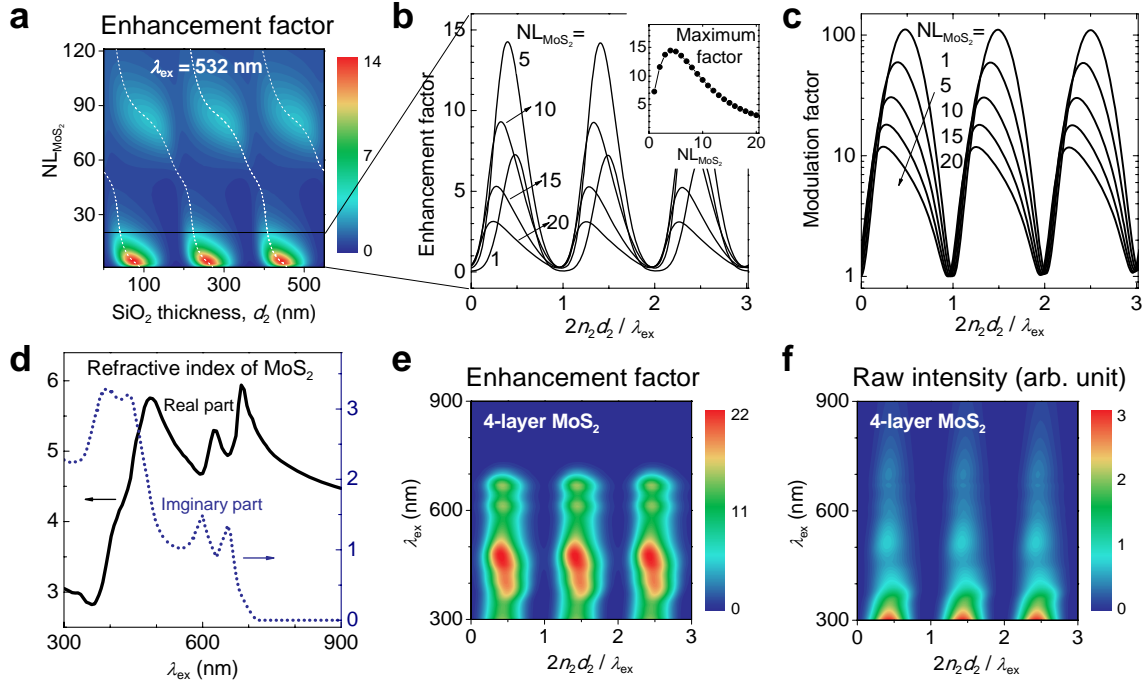

**Figure 5** The dependence of spectral characteristics on three experimental factors: the NL of MoS$_2$ ($NL_{MoS_2}$), the SiO$_2$ thickness ($d_2$) and the excitation wavelength ($\lambda_{ex}$). (a) A contour plot of the calculated enhancement factor as a function of $NL_{MoS_2}$ and $d_2$ at $\lambda_{ex} = 532$ nm. (b) An enlarged plot for low NL regime for typical $NL_{MoS_2}$ values of 1, 5, 10, 15 and 20. (c) Corresponding modulation factor for (b), which is normalized to spectral minimum and reflects the intensity variation due to the change of the SiO$_2$ thickness $d_2$. (d) The highly dispersive refractive index of MoS$_2$. (e) Enhancement factor as a function of $\lambda_{ex}$ and $2n_2d_2/\lambda_{ex}$ for a 4-layer MoS$_2$. The highest value reaches 22. (f) Corresponding raw spectral intensity for (e). The strongest response appears at high-frequency regime due to the well-known quartic relation between scattering cross section and excitation frequency.

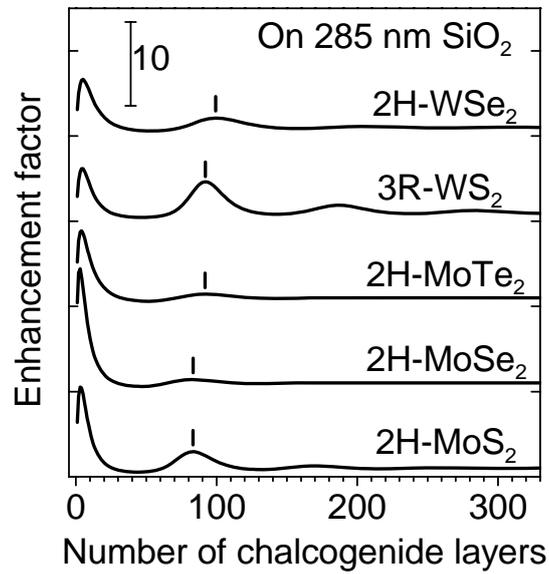

**Figure 6** Calculated enhancement factor for five important layered chalcogenides, $MoS_2$, $MoSe_2$, $MoTe_2$, $WS_2$ and $WSe_2$, on 285-nm $SiO_2$/Si substrates under a 532-nm excitation. The curves are shifted for clarity. The vertical bars indicate the second enhancement peaks for a guide of eyes.

# Quantitative Raman Spectrum and Reliable Thickness Identification for Atomic Layers on Insulating Substrates

By Song-Lin Li,[1,2,*] Hisao Miyazaki,[1] Haisheng Song,[1] Hiromi Kuramochi,[1] Shu Nakaharai,[3] and Kazuhito Tsukagoshi[1,*]

[1]WPI Center for Materials Nanoarchitechtonics (WPI-MANA) and [2]International Center for Young Scientist (ICYS), National Institute for Materials Science, Tsukuba, Ibaraki 305-0044, Japan
[3]Collaborative Research Team Green Nanoelectronics Center, National Institute of Advanced Industrial Science and Technology, Tsukuba, Ibaraki 305-8569, Japan

E-mail: li.songlin@nims.go.jp and tsukagoshi.kazuhito@nims.go.jp

**Content**

1. Expression derivation for interference light

2. Raman response of graphene/graphite

3. $MoS_2$ spectrum with objective lens of varied NA values

4. Calculated spectral intensity for other four chalcogenides

5. Calculated values of intensity ratios for NL identification

6. Refractive indices used in calculations for the four involved optical media

1. Derivation of the expressions for interference light

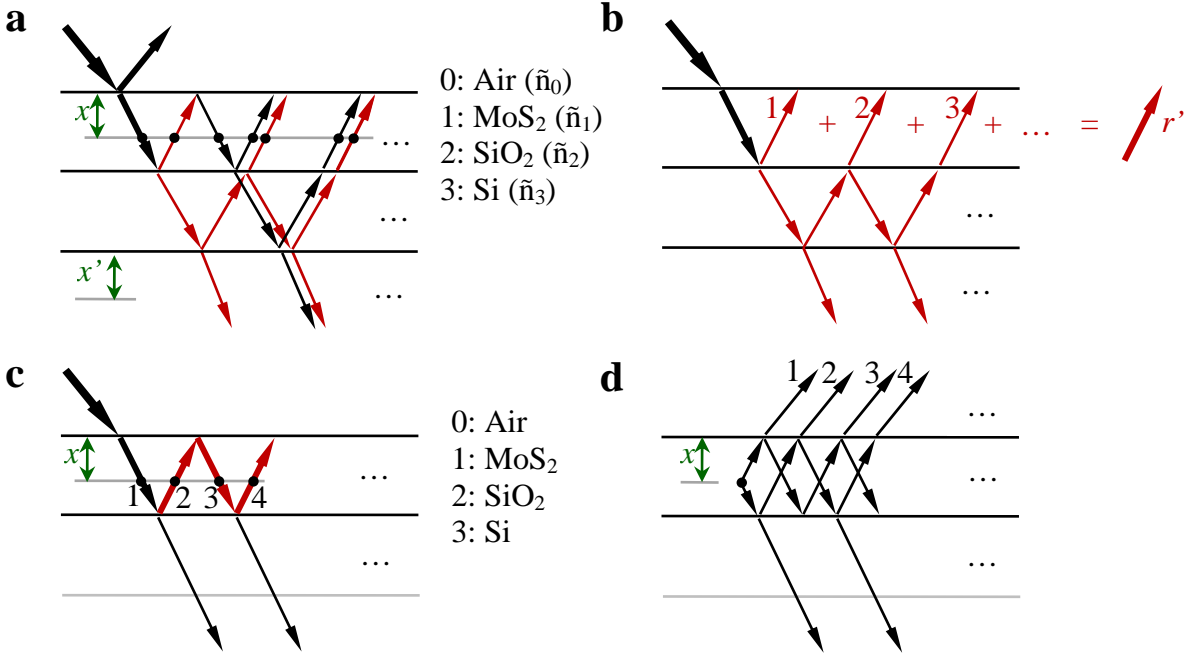

**Figure S1** Schematics of the optical paths in different cases. (a) Overall optical paths for the incident excitation laser light in the four media: air, MoS$_2$, SiO$_2$ and Si substrate. (b) Calculation of the effective reflection at the MoS$_2$/SiO$_2$ interface, including multiple reflections in the lower SiO$_2$ layer. (c), (d) Calculation of the amplitudes of the excitation and scattering light at depth $x$ in MoS$_2$.

For convenience, the four media are assigned an index $i$, and the corresponding (complex) refractive indices are designated by $\tilde{n}_i$, $i$ = 0, 1, 2 and 3 for air, MoS$_2$, SiO$_2$ and Si, respectively. When light propagates through multiple media, the Fresnel transmittance and reflection coefficients at the interfaces are highly dependent on the propagation direction. Specifically, $t_{ij} = 2\tilde{n}_i/(\tilde{n}_i + \tilde{n}_j)$ and $r_{ij} = (\tilde{n}_i - \tilde{n}_j)/(\tilde{n}_i + \tilde{n}_j)$ as a beam reaches the interface $ij$ of media $i$ and $j$ and propagates from medium $i$ to $j$. The two reflection coefficients $r_{ij}$ and $r_{ji}$ from the two sides of the interface $ij$ satisfy the relationship $r_{ij} = -r_{ji}$. There is another relation, $t_{ij}t_{ji} - r_{ij}r_{ji} = 1$, due to the optical reversibility principle. We also define the phase factors $\beta_x = 2\pi\tilde{n}_1 x/\lambda$ and $\beta_j = 2\pi\tilde{n}_j d_j/\lambda$ ($j = 1$ or $2$), representing the phase differences through path $x$ in MoS$_2$ and the whole medium $j$, respectively, where $d_j$ is the thickness of medium $j$ and $\lambda$ is the wavelength of the excitation or scattering light.

1.1 The effective reflection coefficient at the MoS$_2$/SiO$_2$ interface including the lower SiO$_2$ layer
The schematic is shown in Figure S1b and the individual components can be written as
$$b_1 = r_{12}$$
$$b_2 = t_{12} \cdot e^{-i\beta_2} \cdot r_{23} \cdot e^{-i\beta_2} \cdot t_{21} = t_{12}t_{21}r_{23}e^{-2i\beta_2}$$
$$b_3 = t_{12} \cdot e^{-i\beta_2} \cdot r_{23} \cdot e^{-i\beta_2} \cdot r_{21} \cdot e^{-i\beta_2} \cdot r_{23} \cdot e^{-i\beta_2} \cdot t_{21} = b_2 \cdot \left(r_{21}r_{23}e^{-2i\beta_2}\right)^1$$
…
$$b_n = b_2 \cdot \left(r_{21}r_{23}e^{-2i\beta_2}\right)^{n-2}$$

Thus, the total amplitude is

$$r' = r_{12} + t_{12}t_{21}r_{23}e^{-2i\beta_2} \cdot \sum_{n=0}^{\infty}\left(r_{21}r_{23}e^{-2i\beta_2}\right)^n = r_{12} + \frac{t_{12}t_{21}r_{23}e^{-2i\beta_2}}{1-r_{21}r_{23}e^{-2i\beta_2}}.$$

Applying the relationships $r_{21} = -r_{12}$ and $t_{12}t_{21} = 1 - r_{12}^2$ and changing the up-going indices $r_{i+1,i}$ to bottom-going ones $r_{i,i+1}$,

$$r' = r_{12} + \frac{(1-r_{12}^2)r_{23}e^{-2i\beta_2}}{1+r_{12}r_{23}e^{-2i\beta_2}} = \frac{r_{12} + r_{23}e^{-2i\beta_2}}{1+r_{12}r_{23}e^{-2i\beta_2}} \tag{1}$$

1.2 The total amplitude of excitation light at depth $x$ in MoS$_2$
The schematic is shown in Figure S1c and the individual components are expressed as

$$c_1 = t_{01} \cdot e^{-i\beta_x}$$
$$c_2 = t_{01} \cdot e^{-i\beta_1} \cdot r' \cdot e^{-i(\beta_1-\beta_x)} = t_{01}r'e^{-i(2\beta_1-\beta_x)}$$
$$c_3 = t_{01} \cdot e^{-i\beta_1} \cdot r' \cdot e^{-i\beta_1} \cdot r_{10} \cdot e^{-i\beta_x} = c_1 \cdot \left(r'r_{10}e^{-2i\beta_1}\right)^1$$
$$c_4 = t_{01} \cdot e^{-i\beta_1} \cdot r' \cdot e^{-i\beta_1} \cdot r_{10} \cdot e^{-i\beta_1} r' \cdot e^{-i(\beta_1-\beta_x)} = c_2 \cdot \left(r'r_{10}e^{-2i\beta_1}\right)^1$$
$$\ldots$$
$$c_{2n+1} = c_1 \cdot \left(r'r_{10}e^{-2i\beta_1}\right)^n$$
$$c_{2n+2} = c_2 \cdot \left(r'r_{10}e^{-2i\beta_1}\right)^n$$

Thus, the total amplitude of the excitation light at depth $x$ in MoS$_2$ is

$$F_{ex}(x) = \sum_{n=0}^{\infty}\left(t_{01}e^{-i\beta_x} \cdot \left(r'r_{10}e^{-2i\beta_1}\right)^n + t_{01}r'e^{-i(2\beta_1-\beta_x)} \cdot \left(r'r_{10}e^{-2i\beta_1}\right)^n\right) = t_{01} \cdot \frac{e^{-i\beta_x} + r'e^{-i(2\beta_1-\beta_x)}}{1+r'r_{01}e^{-2i\beta_1}} \tag{2}$$

1.3 The amplitude of Raman scattering light from depth $x$ in MoS$_2$
The schematic is shown in Figure S1d and the individual components are

$$d_1 = e^{-i\beta_x} \cdot t_{10} = t_{10}e^{-i\beta_x}$$
$$d_2 = e^{-i(\beta_1-\beta_x)} \cdot r' \cdot e^{-i\beta_1} \cdot t_{10} = t_{10}r'e^{-i(2\beta_1-\beta_x)}$$
$$d_3 = e^{-i\beta_x} \cdot r_{10} \cdot e^{-i\beta_1} \cdot r' \cdot e^{-i\beta_1} \cdot t_{10} = d_1 \cdot \left(r'r_{10}e^{-2i\beta_1}\right)^1$$
$$d_4 = e^{-i(\beta_1-\beta_x)} \cdot r' \cdot e^{-i\beta_1} \cdot r_{10} \cdot e^{-i\beta_1} \cdot r' \cdot e^{-i\beta_1} \cdot t_{10} = d_2 \cdot \left(r'r_{10}e^{-2i\beta_1}\right)^1$$
$$\ldots$$
$$d_{2n+1} = d_1 \cdot \left(r'r_{10}e^{-2i\beta_1}\right)^n$$
$$d_{2n+2} = d_2 \cdot \left(r'r_{10}e^{-2i\beta_1}\right)^n$$

Thus, the amplitude of Raman scattering light from depth $x$ in MoS$_2$ is

$$F_{sc}(x) = \sum_{n=0}^{\infty}\left(t_{10}e^{-i\beta_x} \cdot \left(r'r_{10}e^{-2i\beta_1}\right)^n + t_{10}r'e^{-i(2\beta_1-\beta_x)} \cdot \left(r'r_{10}e^{-2i\beta_1}\right)^n\right) = t_{10} \cdot \frac{e^{-i\beta_x} + r'e^{-i(2\beta_1-\beta_x)}}{1+r'r_{01}e^{-2i\beta_1}} \tag{3}$$

All of the parameters (refractive indices $\tilde{n}_i$ and wavelength $\lambda$) in Equation 3 are for the scattering light rather than the excitation light, which differ from those used in Equation 2. Thus, the fraction terms in Equations 2 and 3, despite having the same form, would lead to different results.

The intensity of output Raman light for the MoS$_2$ flakes is accordingly given by

$$I_{\text{MoS}_2} = \int_0^{d1}\left|F_{ex}(x)F_{sc}(x)\right|^2 dx \tag{4}$$

1.4 The intensity of Raman light from the Si substrate

The thickness of the Si substrate is 0.3 mm. It is treated as a half-infinite media and only bottom-going light exists. The transmittance coefficient at the interface of SiO$_2$ and Si is

$$t_{03}(\lambda_{ex}) = \frac{t_{01}t_{12}r_{23}e^{-i(\beta_1+\beta_2)}}{1 + r_{01}r_{12}e^{-2i\beta_1} + r_{12}r_{23}e^{-2i\beta_2} + r_{01}r_{23}e^{-2i(\beta_1+\beta_2)}}$$

We define a new phase factor $\beta_{x'}(\lambda) = 2\pi\tilde{n}_3 x'/\lambda$ for the Si substrate. Then, the amplitude of excitation light at depth $x'$ in the substrate can be expressed as $t_{03}(\lambda_{ex})e^{-i\beta_{x'}(\lambda_{ex})}$. Similarly, the transmittance coefficient from Si to air is

$$t_{30}(\lambda_{sc}) = \frac{t_{32}t_{21}r_{10}e^{-i(\beta_1+\beta_2)}}{1 + r_{32}r_{21}e^{-2i\beta_2} + r_{21}r_{10}e^{-2i\beta_1} + r_{32}r_{10}e^{-2i(\beta_1+\beta_2)}}$$

and the amplitude of scattering light from depth $x'$ is $t_{30}(\lambda_{sc})e^{-i\beta_{x'}(\lambda_{sc})}$. The two phase factors $\beta_1$ and $\beta_2$ in $t_{03}(\lambda_{ex})$ and $t_{30}(\lambda_{sc})$ are also different due to the change in wavelengths.

The intensity of output Raman light for the Si substrate is accordingly given by

$$I_{Si} = \int_0^\infty \left| t_{03}(\lambda_{ex})e^{-i\beta_{x'}(\lambda_{ex})} t_{30}(\lambda_{sc})e^{-i\beta_{x'}(\lambda_{sc})} \right|^2 dx' \tag{5}$$

2. The Raman response of graphene/graphite on SiO$_2$/Si

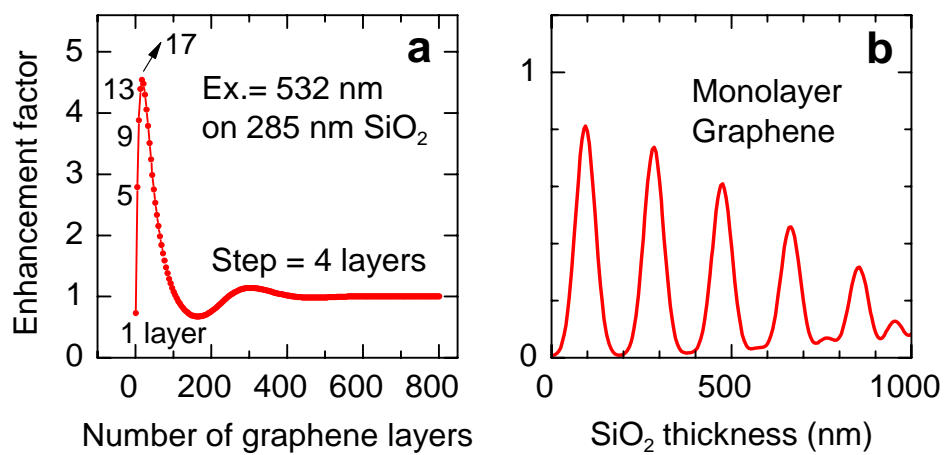

**Figure S2** The calculated enhancement factor as a function of (a) the number of graphene/graphite layers and (b) the SiO$_2$ thickness. A weak second-order IERS peak is predicted in (a).

3. MoS$_2$ spectrum with objective lens of varied NA values

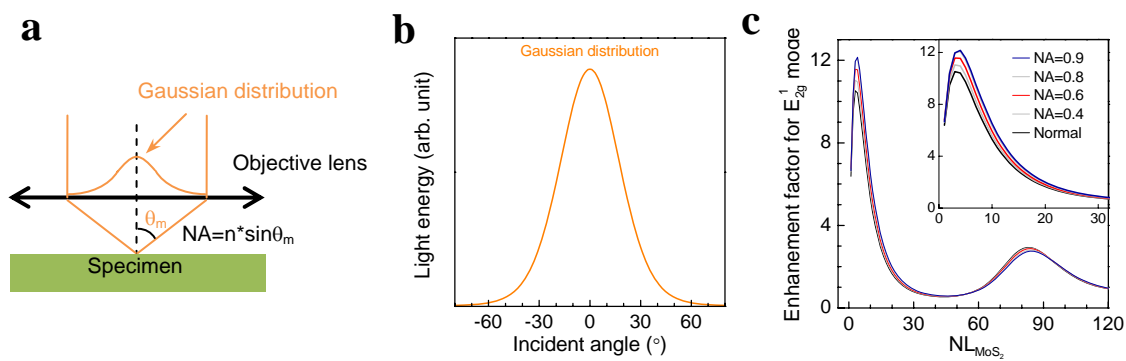

**Figure S3** (a) Schematics for an objective lens. (b) Energy distribution of the excitation laser from our objective lens with a numerical aperture (NA) of 0.9. (c) Calculated enhancement factors of MoS$_2$ spectra for various NA values of objective lenses. No considerable changes are seen in the spectral data when NA values are changed. Therefore, to the first-order approximation the normal incidence assumption is accurate enough.

4. The calculated Raman response of other four chalcogenides

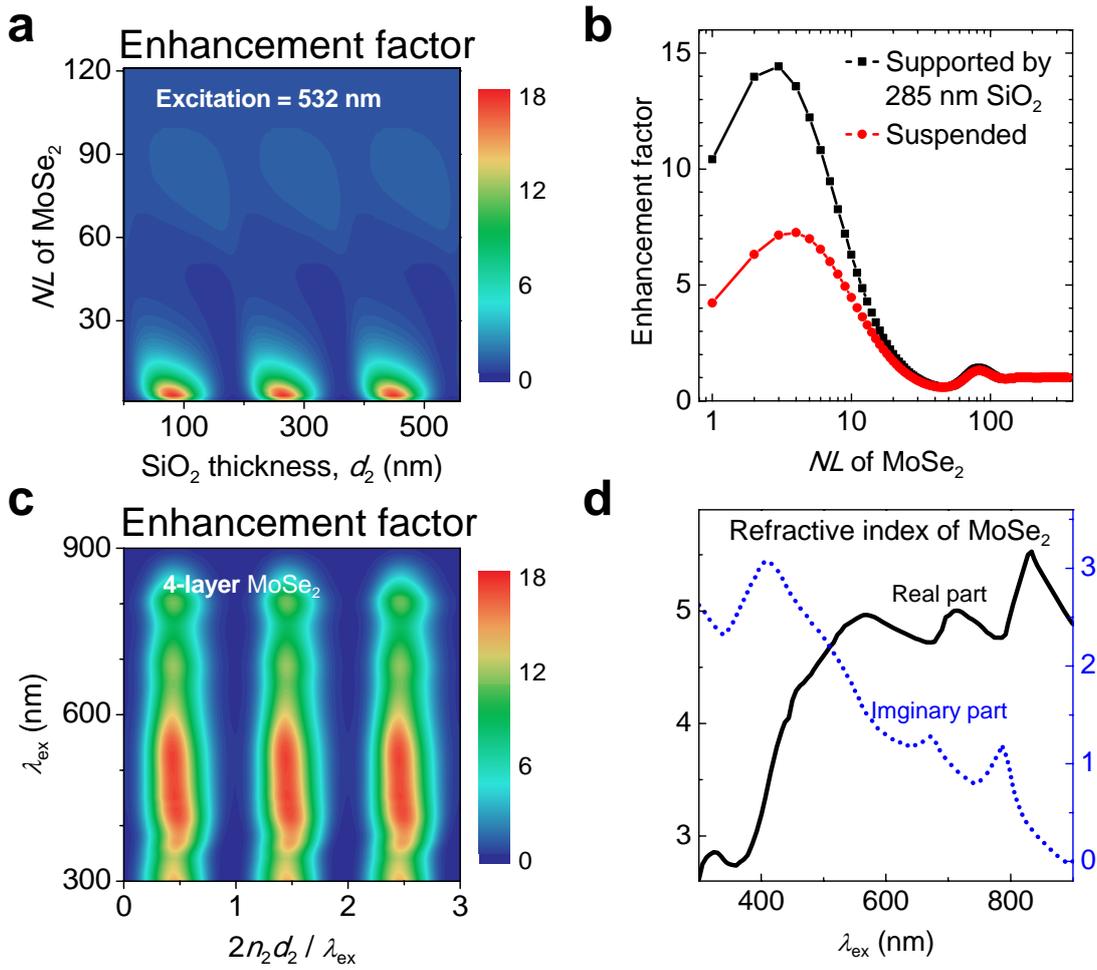

**Figure S4** (a) A contour plot of the calculated enhancement factor of the $E_{2g}^1$ mode (283 cm$^{-1}$) as a function of the MoSe$_2$ $N_L$ and SiO$_2$ thickness ($d_2$) at an excitation wavelength of 532 nm. (b) The enhancement factor for corresponding suspended and supported flakes. (c) The enhancement factor as a function of excitation wavelength ($\lambda_{ex}$) and phase factor ($2n_2d_2\lambda_{ex}$) for a 4-layer flake. (d) The wavelength-dependent refractive index of MoSe$_2$ used in the calculations.

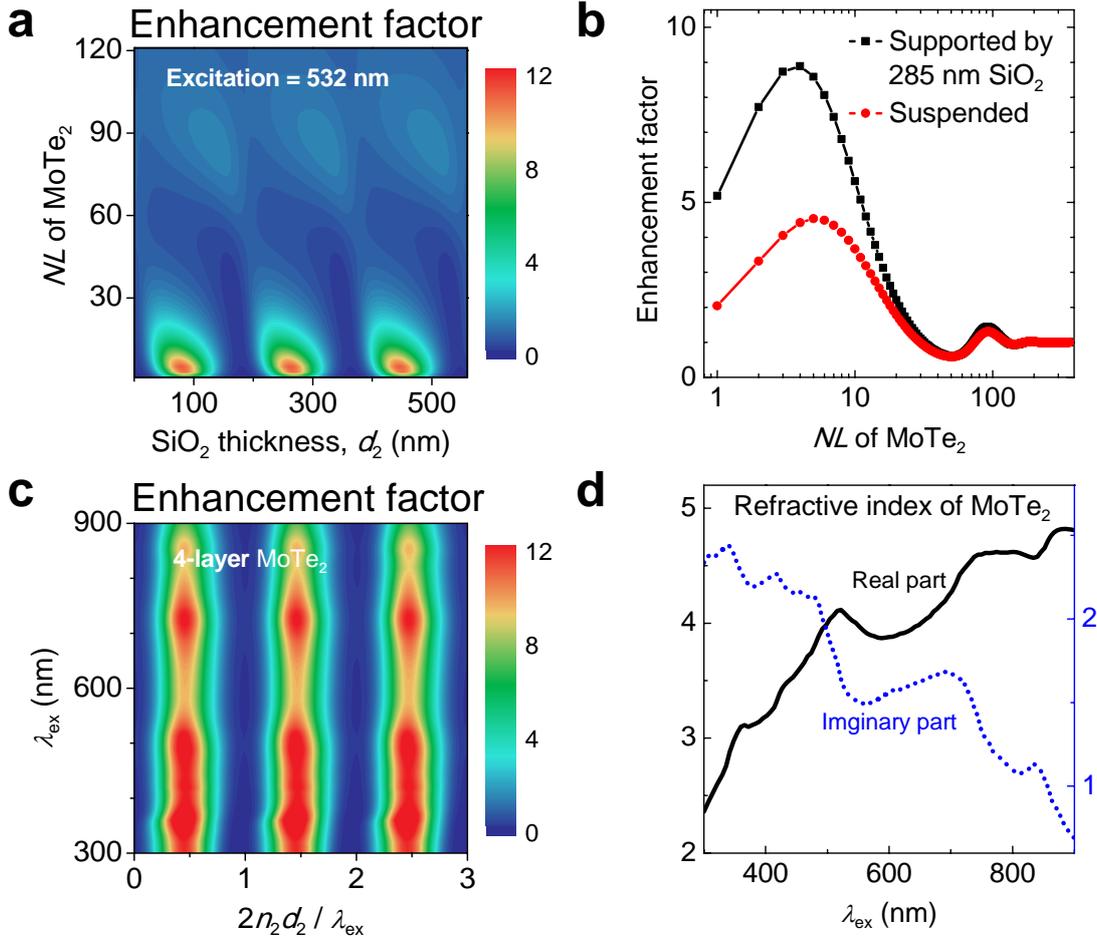

**Figure S5** (a) A contour plot of the calculated enhancement factor of the $E_{2g}^1$ mode (234 cm$^{-1}$) as a function of the MoTe$_2$ $N_L$ and SiO$_2$ thickness ($d_2$) at an excitation wavelength of 532 nm. (b) The enhancement factor for corresponding suspended and supported flakes. (c) The enhancement factor as a function of excitation wavelength ($\lambda_{ex}$) and phase factor ($2n_2d_2\lambda_{ex}$) for a 4-layer flake. (d) The wavelength-dependent refractive index of MoTe$_2$ used in the calculations.

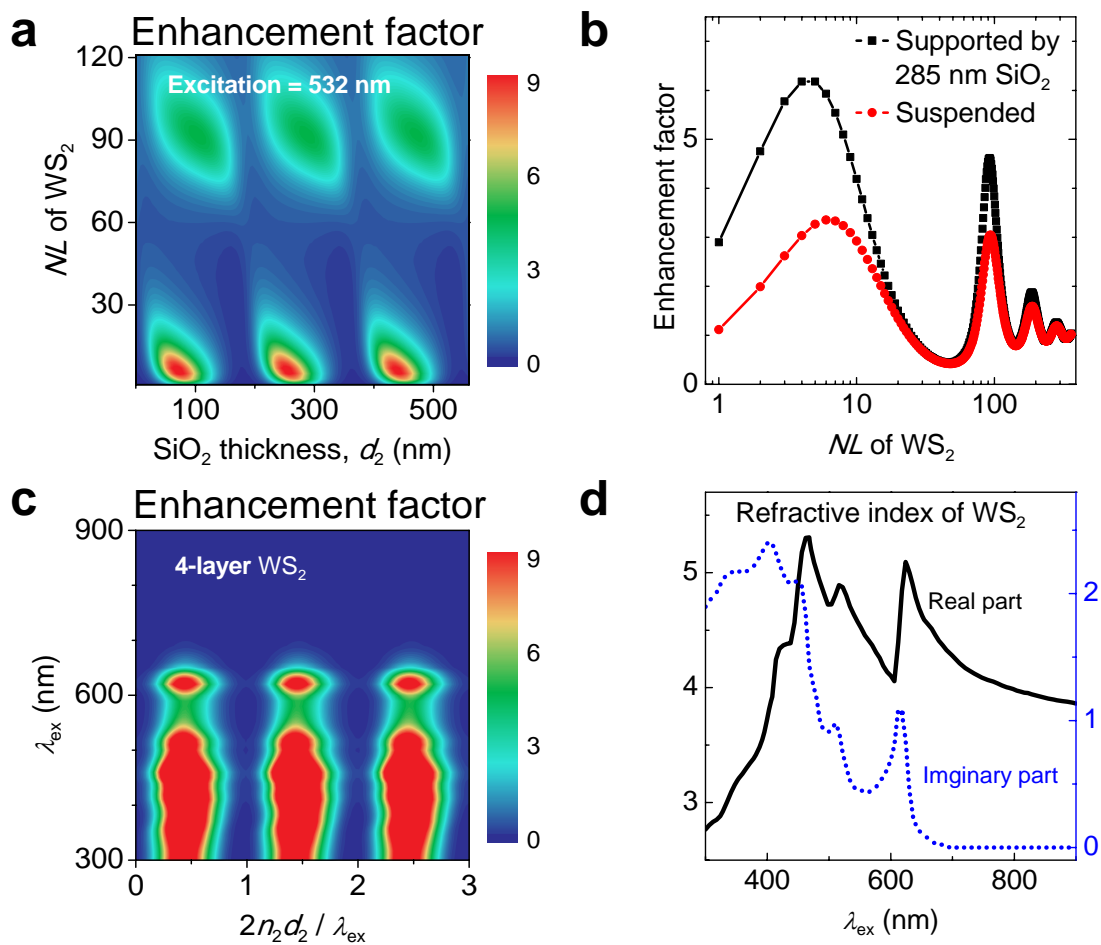

**Figure S6** (a) A contour plot of the calculated enhancement factor of the $E_{2g}^1$ mode (357 cm$^{-1}$) as a function of the WS$_2$ $N_L$ and SiO$_2$ thickness ($d_2$) at an excitation wavelength of 532 nm. (b) The enhancement factor for corresponding suspended and supported flakes. (c) The enhancement factor as a function of excitation wavelength ($\lambda_{ex}$) and phase factor ($2n_2 d_2 \lambda_{ex}$) for a 4-layer flake. (d) The wavelength-dependent refractive index of WS$_2$ used in the calculations.

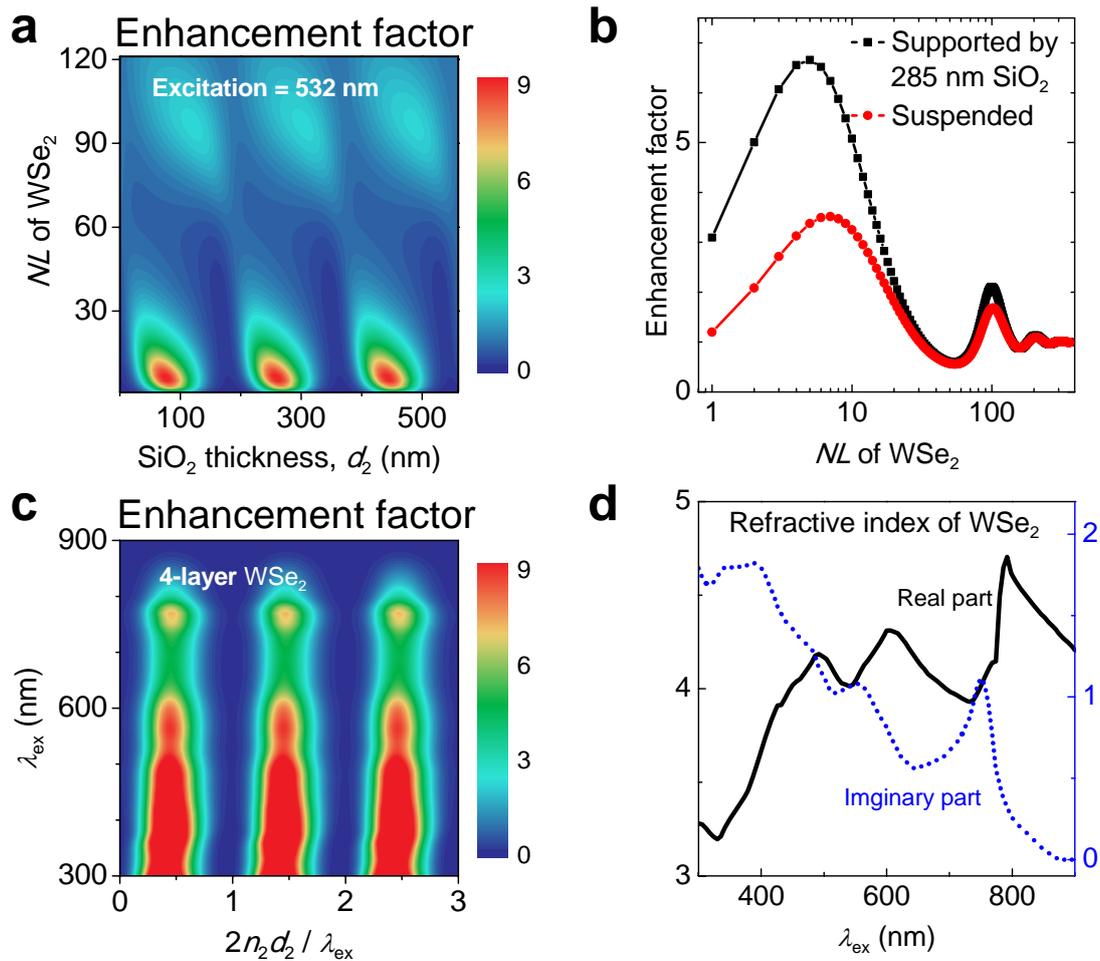

**Figure S7** (a) A contour plot of the calculated enhancement factor of the $E_{2g}^1$ mode (247 cm$^{-1}$) as a function of the WSe$_2$ $N_L$ and SiO$_2$ thickness ($d_2$) at an excitation wavelength of 532 nm. (b) The enhancement factor for corresponding suspended and supported flakes. (c) The enhancement factor as a function of excitation wavelength ($\lambda_{ex}$) and phase factor ($2n_2d_2\lambda_{ex}$) for a 4-layer flake. (d) The wavelength-dependent refractive index of WSe$_2$ used in the calculations.

5. Calculated values of intensity ratios for NL identification

**Table S1** The values for the area ratios of the MoS$_2$ $E^1_{2g}$ to the 520-cm$^{-1}$ Si modes.

| Number of MoS$_2$ layers | 1 | 2 | 3 | 4 | 5 | 6 | 7 | 8 | 9 | 10 |
|---|---|---|---|---|---|---|---|---|---|---|
| Intensity ratio of $E^1_{2g}$ to Si | 0.3 | 0.5 | 0.8 | 1.0 | 1.3 | 1.5 | 1.8 | 2.0 | 2.2 | 2.4 |
| Intensity ratio of A$_{1g}$ to Si | 0.5 | 0.9 | 1.4 | 1.8 | 2.2 | 2.7 | 3.1 | 3.4 | 3.8 | 4.1 |
| Number of MoS$_2$ layers | 11 | 12 | 13 | 14 | 15 | 16 | 17 | 18 | 19 | 20 |
| Intensity ratio of $E^1_{2g}$ to Si | 2.6 | 2.7 | 2.9 | 3.0 | 3.2 | 3.3 | 3.4 | 3.5 | 3.6 | 3.7 |
| Intensity ratio of A$_{1g}$ to Si | 4.5 | 4.8 | 5.0 | 5.3 | 5.6 | 5.8 | 6.0 | 6.2 | 6.3 | 6.5 |

6. Refractive indices of the four involved optical media

**Table S2** A list of refractive indices used for calculating the $E^1_{2g}$ mode for five layered chalcogenides. All of the values were adopted from Refs. 35–38 in the manuscript.

|  | Raman shift of the $E^1_{2g}$ mode (cm$^{-1}$) | Wavelength (nm) | Air | Chalcogenide | SiO$_2$ | Si substrate |
|---|---|---|---|---|---|---|
| MoS$_2$ | - | 532.3 | 1 | 5.211-1.128i | 1.461 | 4.149-0.0426i |
|  | 383.4 | 543.4 | 1 | 5.084-1.068i | 1.460 | 4.108-0.0442i |
| MoSe$_2$ | - | 532.3 | 1 | 4.859-2.056i | 1.461 | 4.149-0.0426i |
|  | 283 | 540.4 | 1 | 4.892-1.976i | 1.460 | 4.118-0.0470i |
| MoTe$_2$ | - | 532.3 | 1 | 4.065-1.602i | 1.461 | 4.149-0.0426i |
|  | 234 | 539.0 | 1 | 4.025-1.554i | 1.460 | 4.123-0.0480i |
| WS$_2$ | - | 532.3 | 1 | 4.726-0.737i | 1.461 | 4.149-0.0426i |
|  | 357 | 542.6 | 1 | 4.604-0.552i | 1.460 | 4.110-0.0448i |
| WSe$_2$ | - | 532.3 | 1 | 4.024-1.032i | 1.461 | 4.149-0.0426i |
|  | 247 | 539.4 | 1 | 4.015-1.046i | 1.460 | 4.122-0.0478i |